\documentclass[12pt]{article}
\usepackage{latexsym,amsmath,amssymb,amsbsy,graphicx}
\usepackage[cp1251]{inputenc}
\usepackage[T2A]{fontenc}
\usepackage[russian,english]{babel}
\usepackage[space]{cite} 

\begin{document}

\centerline {{\bf The histeresis of the indices of solar activity
and of the ionospheric}} \centerline {{\bf indices in 11-yr cycles.
The histeresis of the stellar activity indices}} \centerline {{\bf
in the cyclic activity similar to the Sun}}

\bigskip

\centerline {E.A.Bruevich $^{a}$, G.V. Yakunina$^{a}$,}

\centerline {T.V. Kazachevskaya $^{b}$, V.V. Katyushina $^{b}$, A.A.
Nusinov $^{b}$}

\bigskip
\bigskip
\centerline {{$^a${Sternberg Astronomical Institute, MSU, Moscow,
Russia }}}
\centerline {$^b${Fedorov Institute of Applied
Geophysics, Moscow, Russia}}

\centerline {E-mail: $^a${red-field@yandex.ru},
$^b${nusinov@ipg.geospace.ru}}

{\bf Abstract.} The effects of hysteresis, which manifests itself
in an ambiguous relationship between different indices for solar
activity on the phases of rise and decline in the cycle, are
analyzed for the indices of solar activity (which characterize the
out coming radiation of the solar photosphere, chromosphere and
corona) and also for the ionospheric indices. In the cycles 21 -
23 which are significantly different in amplitude, the effect of
hysteresis manifests itself to a variable extent. The stars with
the well-determined cycles were examined to detect the effect of
hysteresis between the chromosphere's S-index CaII fluxes versus
the photosphere's fluxes $F_{photosphere}$ .

\bigskip

{\it Key words.} Sun: activity indices, solar-type stars: the HK-
project, activity cycles.
\bigskip

\vskip12pt
\section{Introduction}
\vskip12pt

In solar and solar-terrestrial physics the phenomenon of
hysteresis has long been known [1, 2, 3]. This effect manifests
itself (for different levels of solar activity) in an ambiguous
dependence of the radiation of the solar photosphere, chromosphere
and corona, as well as indices characterizing the state of Earth's
ionosphere. We have chosen to study the effect of the hysteresis
in solar activity versus off the index $F_{10.7}$ (radio flux on
the wave 10.7 cm -- 2800 MHz).

The aim of this work is to study the variations of indices of
activity during three solar cycles: (1) the changes in the values
of the activity indices (AI - Activity Indices) depending on the
solar activity level (i.e. versus the $F_{10.7}$ index), and (2)
the analysis of the ambiguous dependence of the relationship of
indices of activity versus $F_{10.7}$ on the rise and decline
phases of the solar cycle.

For the monthly values of 7 activity indices AI we have determined
the coefficients of the quadratic regression for the correlations
with an  $F_{10.7}$ index (AI $\leftrightarrow $ $F_{10.7}$). Also
we determined the coefficients of a quadratic regression for 5
daily values of indices of solar and ionospheric activity (AI
$\leftrightarrow $ $F_{10.7}$).

Thus a series of observations of the indices of the SSN, TSI and
$F_{L \alpha}$ were analyzed for monthly and daily data.

We also analyzed the hysteresis of the solar-type stars (with a
stable cycles of activity similar to the sun's cycles) between the
radiation fluxes from their photospheres and chromospheres (in the
lines H and K CaII) from the Mount Wilson HK-project and Lowell
observatory data [4,5].

In this study we used the data of observations of $F_{10.7}$, SSN
and other indices of solar and ionospheric activity from the
archives of the NOAA National Geophysical Data Center and of the
Solar-Geophysical Data Reports [6, 7].

\section{The indices of activity at different phases of the 11-yr solar cycle.
The hysteresis of their monthly values}

We study the hysteresis of the solar and ionospheric indices,
varying in the 11-yr cycle, depending on the overall level of
solar activity which is defined by the value of $F_{10.7}$.

The solar radio microwave flux at wavelengths 10.7 cm $F_{10.7}$ has
also the longest running series of observations started in 1947 in
Ottawa, Canada and maintained to this day at Penticton site in
British Columbia. This radio emission comes from high part of the
chromosphere and low part of the corona. $F_{10.7}$ radio flux  has
two different sources: thermal bremsstrahlung (due to electrons
radiating when changing direction by being deflected by other
charged participles) and gyro-radiation (due to electrons radiating
when changing direction by gyrating around magnetic fields lines).
These mechanisms give rise to enhanced radiation when the
temperature, density and magnetic fields are enhanced. So $F_{10.7}$
is a good measure of general solar activity. $F_{10.7}$ data are
available at http://radbelts.gsfc.nasa.gov. The results of
measurements of the $F_{10.7}$ flux are expressed in units of solar
flux (sfu): $1 sfu = 10^{-22} W \cdot m^{-2} \cdot Hz^{-1}$.

SSN has the longest series of observations (monthly averaged
values of the relative sunspot number are known since 1749, and
annual averaged are known since 1700). The solar activity is
currently studying with the help of other indices (simultaneously
with the SSN), which are carrying the quantitative information
about the general levels of activity in those areas of the solar
atmosphere where they are formed.

\begin{figure}[tbh!]
\centerline{
\includegraphics[width=140mm]{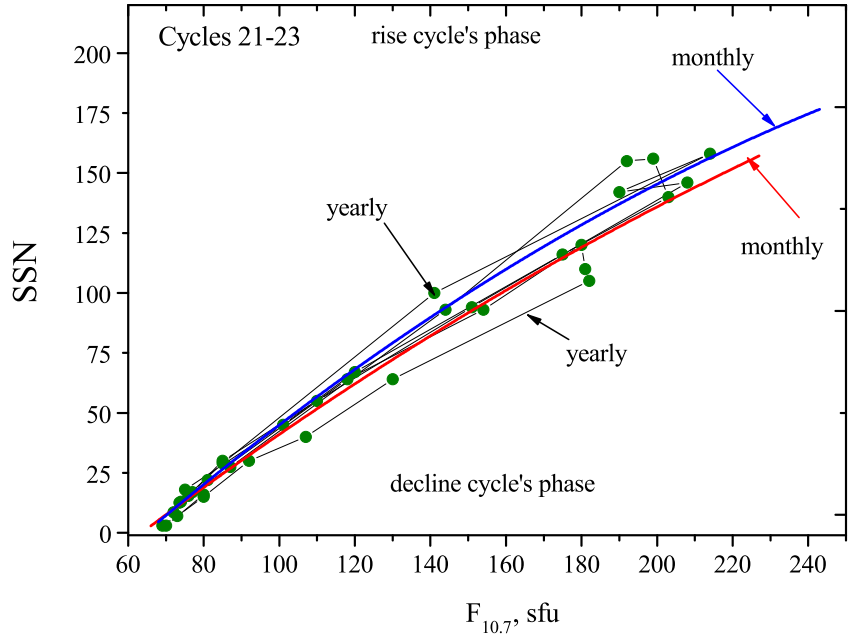}}
 \caption{Monthly and annual averaged SSN depending on the monthly
and annual averaged $F_{10.7}$.} {\label{Fi:Fig1}}
\end{figure}

The radiation flux in the ultraviolet  $H {L \alpha}$ line (121.6
nm), $F_{L \alpha}$ -- an important indicator of solar radiation,
carrying the information of the chromosphere and bottom of the
transition zone, [8].

Ultraviolet flux and X-rays represent a relatively small part in
the total energy flux of solar radiation. This part of the
spectrum is highly variable in time and strongly affects the upper
atmosphere of the Earth causing ionization and dissociation of
atmospheric components, leading to the formation of the
ionosphere. The activity index $F_{L \alpha}$ is measured in
$10^{11}photon \cdot cm^{-2} \cdot sec^-1$. Earlier the  $F_{L
\alpha}$ index was analyzed in our paper [9] on the basis of data
obtained with SUPR devices which was installed on different
Russian satellites.

The Mg II 280 nm is important solar activity indicator of radiation,
derived from daily solar observations of the core-to-wing ratio of
the Mg II doublet at 279,9 nm provides a good measure of the solar
UV variability and can be used as a reliable proxy to model extreme
UV (EUV) variability during the solar cycle [10]. The Mg II
observation data were obtained from several satellite's (NOAA,
ENVISAT) instruments. Comparison of the NOAA and ENVISAT Mg II index
observation data shows that both the MgII indexes agree to within
about 0,5\%. We used both the NOAA and ENVISAT Mg II index observed
data from Solar-Geophysical Data Bulletin [7] and [10]. In [11] it
was showen an extremely good fit between the 30,4 nm emission (the
main component of EUV-emission) and the NOAA Mg II index. In
[11,12]it have been noted the close linear relationship between the
Mg II index and total solar irradiance -- TSI.

TSI is the solar irradiance is the total amount of solar energy at a
given wavelength received at the top of the earth's atmosphere per
unit time. When integrated over all wavelengths, this quantity is
called the total solar irradiance (TSI) previously known as the
solar constant. Regular monitoring of TSI has been carried out since
1978. We use the TSI data set from NGDC web site
http://www.ngdc.noaa.gov and combined observational data from
National Geophysical Data Center Solar and Terrestrial Physics [6].
The importance of UV/EUV influence to TSI variability (Active
Sun/Quiet Sun) was pointed in [13]. There were indicated that up to
63,3 \% of TSI variability is produced at wavelengths below 400 nm.
Towards activity maxima the number of sunspots grows dramatically.
But on average the TSI is increased by about 0,1\% from minimum to
maximum of activity cycle. This is due to the increase amounts of
bright features, faculae and network elements on the solar surface.
The total area of the solar surface covered by such features rises
more strongly as the cycle progresses than the total area of dark
sunspots. Some physics-based models have been developed with using
the combined proxies describing sunspot darkening (sunspot number or
areas) and facular brightening (facular areas, Ca II or Mg II
indices), see [14].

We also analyzed  rapid processes on the Sun: Counts of Flares --
monthly counts of grouped solar flares (according [7] the term
'grouped' means observations of the same event by different sites
were lumped together and counted as one) and Flare Index, summing
the "total energy" emitted in the optical range in flares.

Earlier in [15] for the cycles 21 -23, we investigated the
correlation between several solar activity indices and $F_{10.7}$.
It turned out that the correlation coefficient depends on the phase
of cycles of activity and its dependence on time is asymmetric for
the rise and decline cycle's phases. This fact, along with detected
shifts in time between the maximums of the values of indices of
solar activity during one cycle indicates the existence of
hysteresis of the solar indices.

We have analyzed the data for the rise and decline phases for three
cycles of activity separately. The effect of hysteresis, with some
differences in the values of the coefficients of the quadratic
regression between the indices and $F_{10.7}$ was observed in each
of these cycles.

It should be noted that according to the observations of indices of
activity, in particular, SSN and TSI, the dependence of the ratio of
the observed to the estimated AI in the regression ratios for the
period 1950-1990 has changed markedly since 1990, previously this
ratio was constant, see [16]. This has affected the values of
coefficients of quadratic regressions between indices and $F_{10.7}$
in the cycle 23 compared to the cycles 21 and 22.

\begin{figure}[tbh!]
\centerline{
\includegraphics[width=140mm]{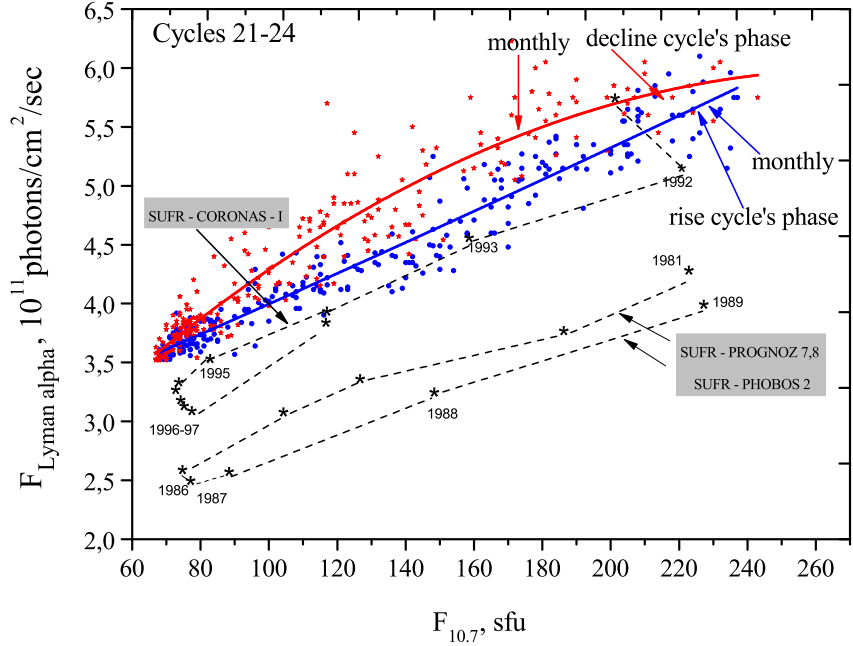}}
 \caption{Monthly averaged  $F_{L\alpha}$ depending on the monthly
averaged $F_{10.7}$ } {\label{Fi:Fig1}}
\end{figure}

Figures 1 - 4 shows the monthly averaged values of SSN,
$F_{L\alpha}$, MgII, TSI, Flare Index and $F_{530.3}$ depending on
$F_{10.7}$ for the cycles 21 -- 23. Due to significant scatter of
data,  it is advisable to compare the regression equations for each
phase (rise and decline) separately. It was showed that to analyze
the observation of solar indices is enough to use with polynomials
of second order.

For the rise cycle's phases and the decline cycle's phases the solid
lines are shown the quadratic regression lines. It is seen that the
effect of hysteresis manifests itself in all activity indices and
reach up to 10 - 15 \% with the exception of the MgII index and the
TSI, which have very small  relative variations in the cycles of
activity (Fig. 3a,b).

Fig. 1 also shows the dependence of the yearly averaged SSN from the
yearly averaged $F_{10.7}$. It can be seen that the hysteresis value
(the maximum relative deviation in the cycle) in yearly averaged
case is increased to 20 \%.

In Fig. 2 the solid lines are the lines of quadratic regression for
the synthetic series of the $F_{L \alpha}$ (Composite Solar Lyman
alpha the data from the Laboratory for Atmospheric and Space Physics
at the University of Colorado, see
http://lasp.colorado.edu/lisird/tss/composite\_lyman\_alpha.html.

 The dashed lines refer to the yearly
averaged data over the year of observations of SUFR (PHOBOS 2,
PROGNOZ 7,8, CORONAS) for the cycles 21 and 22, see [9].

\begin{figure}[h!]
   \centerline{\hspace*{0.005\textwidth}
               \includegraphics[width=70mm]{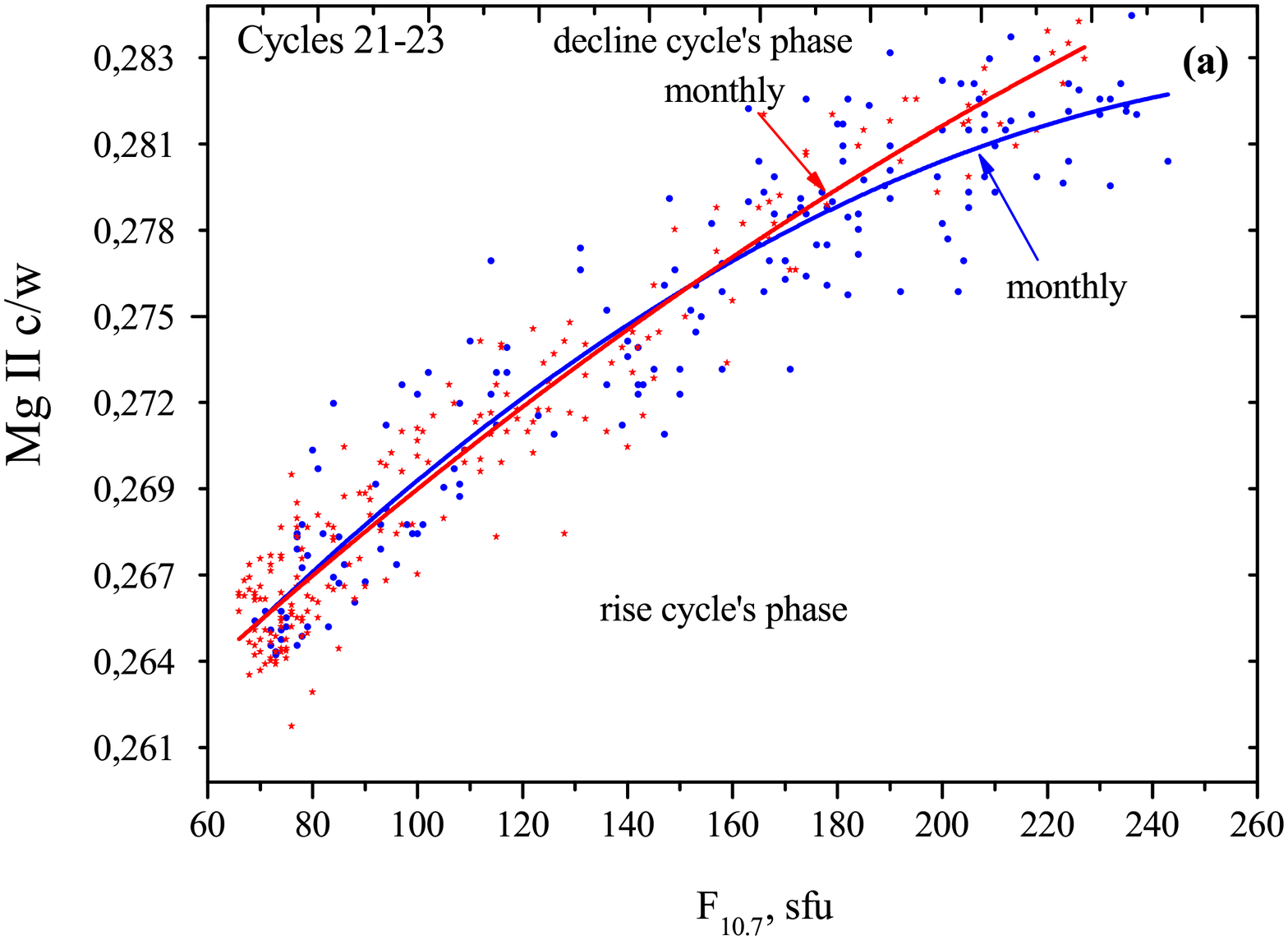}
               \includegraphics[width=70mm]{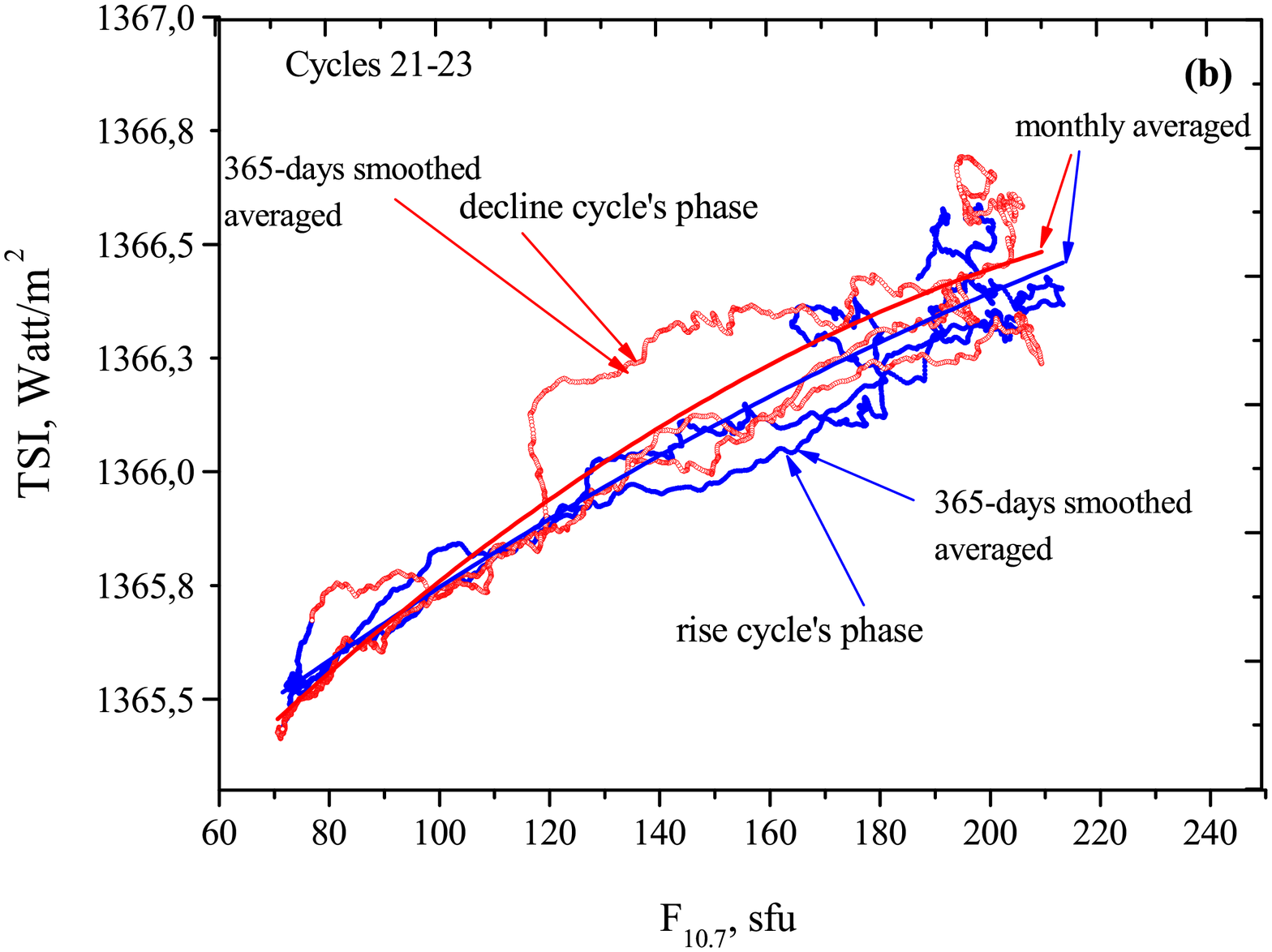}
              }

\caption{The MgII (a) monthly averaged values and the TSI (b)
monthly averaged values versus the monthly averaged $F_{10.7}$}
 {\label{Fi:Fig3}}
\end{figure}

In Fig. 3a,b the solid lines are the lines of quadratic regressions
of  monthly averaged values of the MgII and the TSI in the cycles
21-23. In Fig. 3b we also present the 365-day moving averages of
the TSI for the rise and decline  phases of the cycle 21. We can see
that the effect of hysteresis in the case of the 365-day moving
averages is more pronounced.

Fig. 4a,b shows that the indices associated with the rapidly varying
solar activity are characterized by a hysteresis value of about 10
-- 20 \%.

The quadratic regressions which are shown in Fig. 1 - 4, are
described by equation (1):

\begin{equation}
    AI = A + C1 \cdot F_{10.7} + C2 \cdot F_{10.7}^2 \,.
   \end{equation}

here AI is solar activity index,

A is the intercept of polinomial regression,

C1 and C2 are the coefficients of polinomial regression. These
coefficients are presented in the Table 1.

\begin{figure}[h!]
   \centerline{\hspace*{0.005\textwidth}
               \includegraphics[width=70mm]{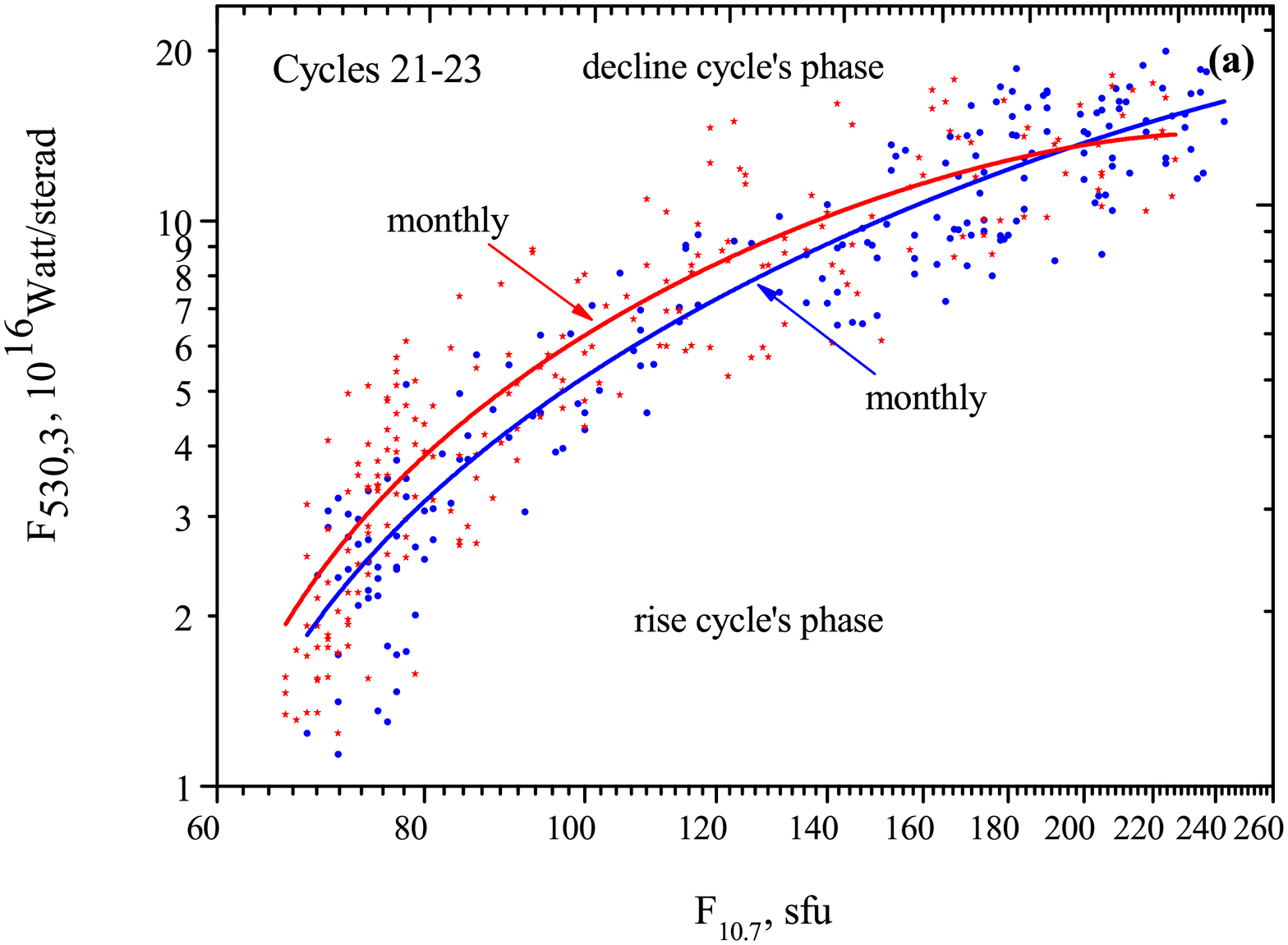}
               \includegraphics[width=70mm]{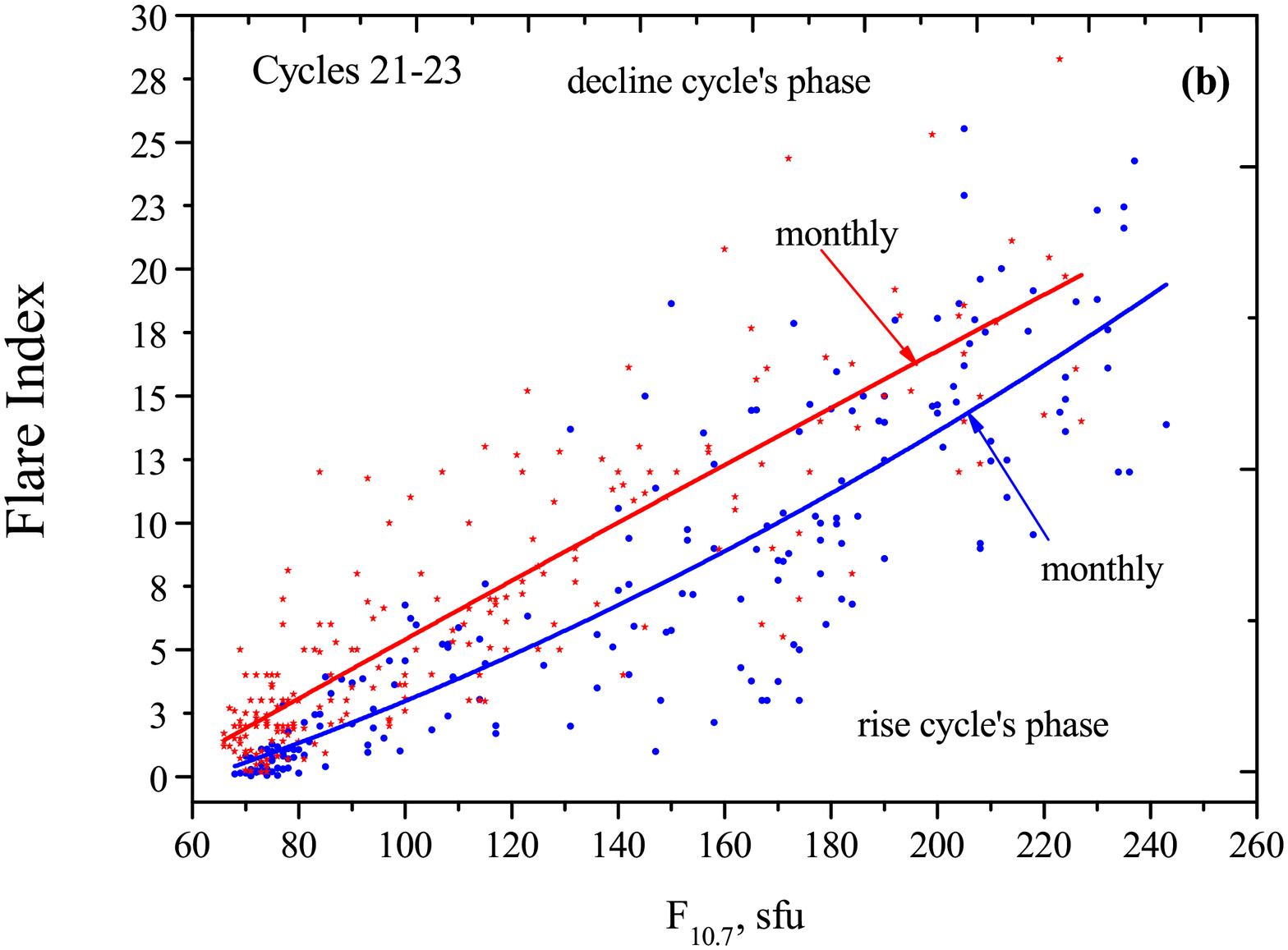}
              }

\caption{The monthly averaged values of the Flare Index (a) and
radiation flux $F_{530.3}$ (b) versus the monthly averaged
$F_{10.7}$}
 {\label{Fi:Fig4}}
\end{figure}

\begin{table}
\caption{The coefficients of the quadratic regression for monthly
averaged values of the activity indices AI in the cycles 21, 22 and
23  in the phases of rise (R) and decline (D)}
\begin{tabular}{ccccccc}     
  \hline                   
AI $\leftrightarrow$ $F_{10.7}$             & A       & C1    & C2    & $\sigma_A$  & $\sigma_{C1}$& $\sigma_{C2}$    \\
  MgII $\leftrightarrow$ $F_{10.7}$,R       & 0.252   & 2.1E-4& 1.9E-5& 0.0014      & 3.7E-7       & 6.7E-8       \\
MgII $\leftrightarrow$ $F_{10.7}$,D         & 0.254   & 1.7E-4& 1.8E-5 & 0.0009     & 1.9E-7       & 5.6E-8       \\
$F_{L \alpha}$$\leftrightarrow$ $F_{10.7}$,R& 2.126  & 0.022 & -2.7E-5 & 0.16       & 0.0025       & 8.5E-6       \\
$F_{L \alpha}$$\leftrightarrow$ $F_{10.7}$,D& 1.543  & 0.035 & -7.4E-5 & 0.18       & 0.0028       & 9.9E-6       \\
SSN $\leftrightarrow$ $F_{10.7}$,R          & -87.7  & 1.41  & -8.8E-4 & 11.9       & 0.198        & 6.7E-8       \\
SSN $\leftrightarrow$ $F_{10.7}$,D          & -79.7  & 1.34  & -0.001  & 6.2        & 0.103        & 5.6E-8       \\
Count Fl$\leftrightarrow$ $F_{10.7}$,R      & -356.5 & 6.8   & -0.01   & 40.17      & 0.68         & 4.7E-3       \\
Count Fl$\leftrightarrow$ $F_{10.7}$,D      & -206.2 & 3.21  &0.006    & 32.48      & 0.35         & 4.9E-4       \\
TSI $\leftrightarrow$ $F_{10.7}$,R          &1364.3  & 0.019 & -4.5E-5 & 0.184      & 0.003        & 8.7E-6       \\
TSI$\leftrightarrow$ $F_{10.7}$,D           & 1364.5 & 0.015 &
-2.6E-5 & 0.161      & 0.002      & 9.2E-6       \\ \hline

\end{tabular}
\end{table}

In the Table 1 the regression coefficients A, C1 and C2 are
indicated for the rise (R) and decline (D) phases of solar cycle.
  $\sigma_A$ , $\sigma_{C1}$, $\sigma_{C2}$  are the standard deviations of the values of the
regression coefficients. The statistical estimates show that the
regression coefficients for the phases of rise and decline in the
cycles of activity vary with the level of significance which is
equal to $\alpha$=0.05.

\section{Hysteresis of daily values of solar and ionospheric activity indices in
different phases of the 11-year cycle.}

Earlier in this paper we have identified effects of hysteresis for
monthly values. It is of interest to perform well as the average
daily values of the indices. The investigated indices of the
activity of the Sun AI are described by the same quadratic
regression (1). The regression coefficients for the daily averaged
values are presented in the Table 2, all the symbols in this Table 2
are corresponded to the designations of Table 1.

Fig. 5 shows that the daily SSN data are characterized by a large
scatter of values, as in the paper [3]. Solid lines indicate the
regression line for the daily data of SSN observations, while the
dashed show the monthly data. It is seen that for daily values, the
effect of hysteresis is more pronounced.

In Fig. 6a,b we also see the large scatter of the daily values of
indices. The solid lines show the regression line for the daily
data, the dashed lines in Fig. 6a show the regression line for
monthly data.

\begin{figure}[tbh!]
\centerline{
\includegraphics[width=140mm]{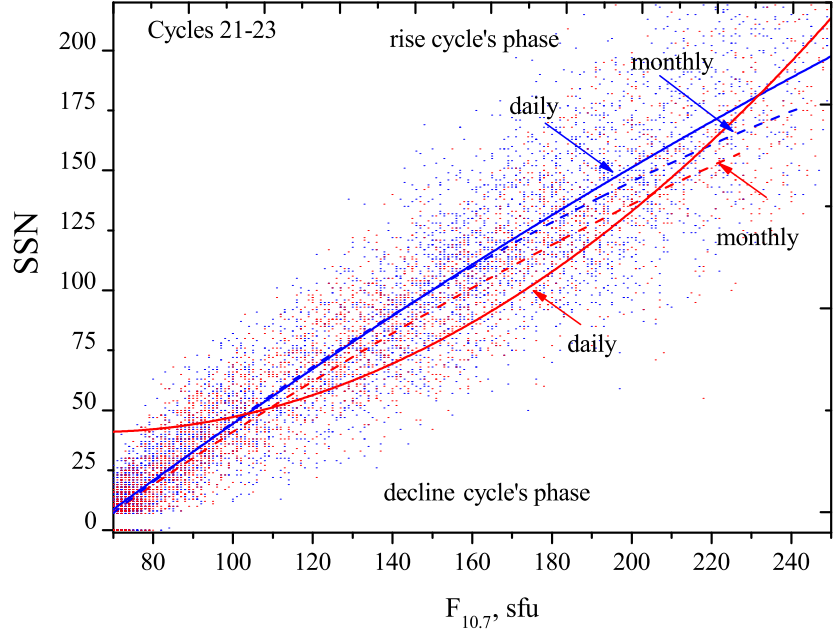}}
 \caption{The daily values of the SSN depending on the daily averaged
$F_{10.7}$  } {\label{Fi:Fig5}}
\end{figure}

\begin{figure}[h!]
   \centerline{\hspace*{0.005\textwidth}
               \includegraphics[width=70mm]{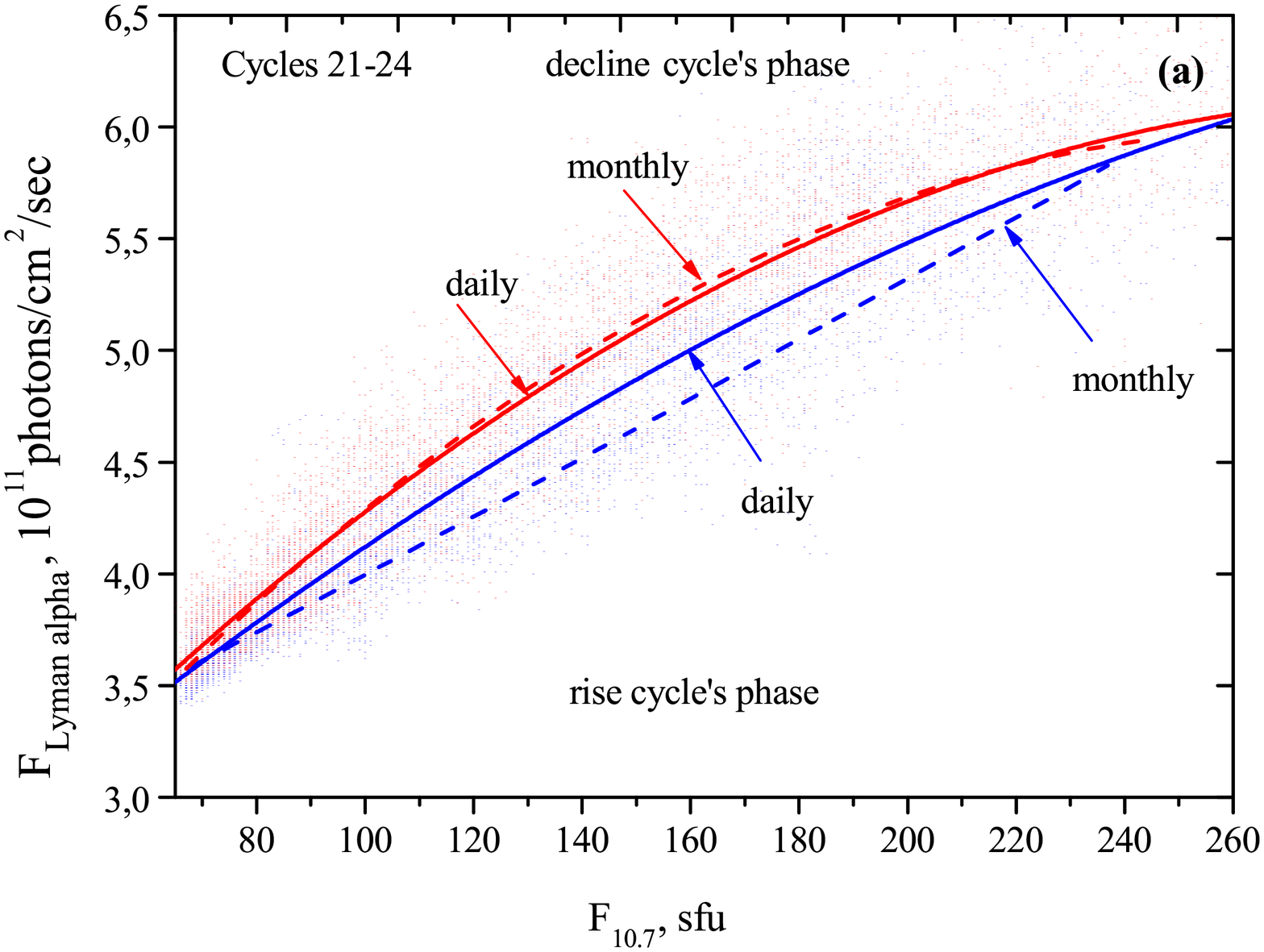}
               \includegraphics[width=70mm]{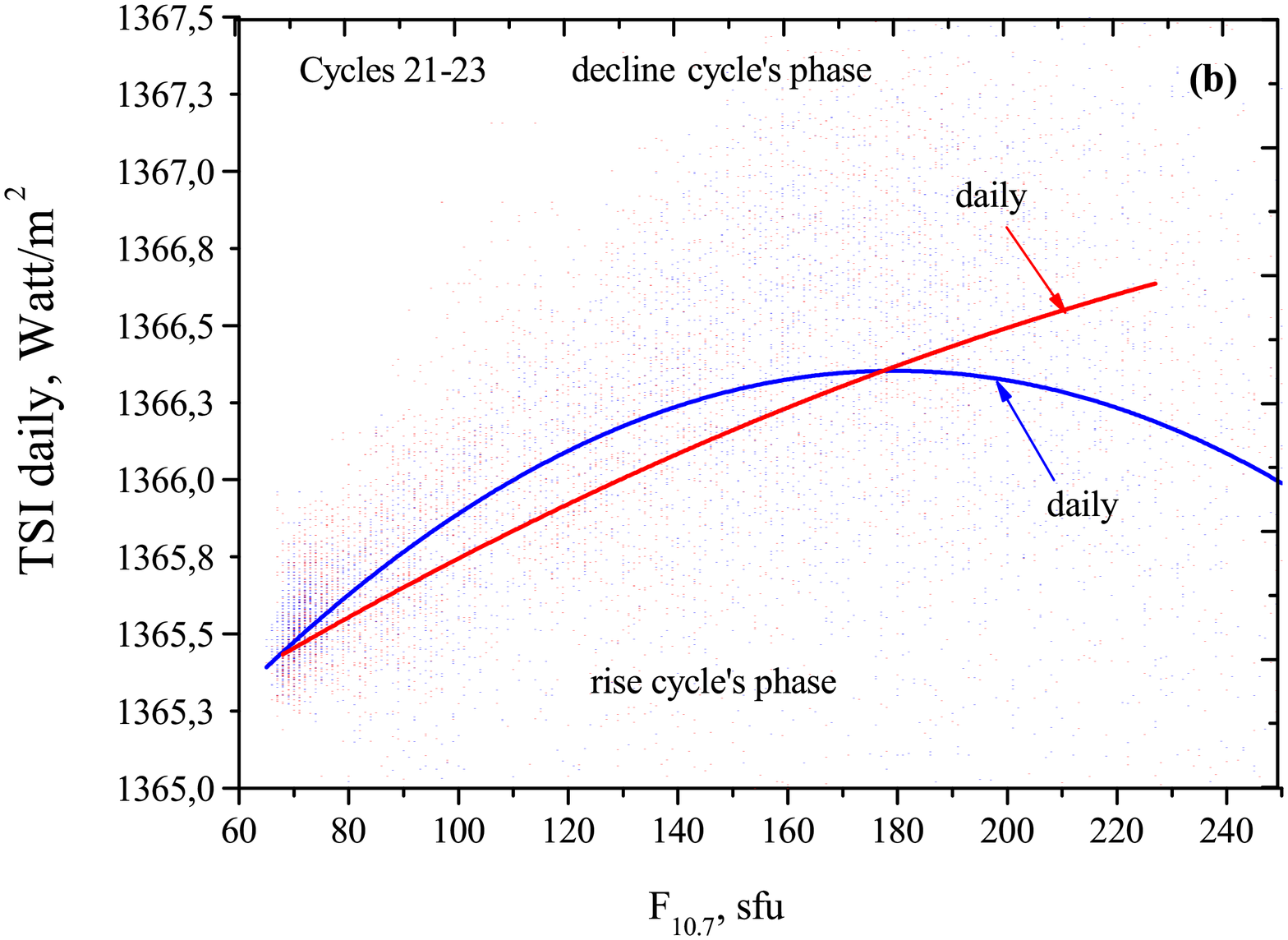}
              }

\caption{The daily values of  the $F_{L\alpha}$ (a) and of  the TSI
(b) depending on the daily averaged $F_{10.7}$}
 {\label{Fi:Fig6}}
\end{figure}

\begin{figure}[h!]
   \centerline{\hspace*{0.005\textwidth}
               \includegraphics[width=70mm]{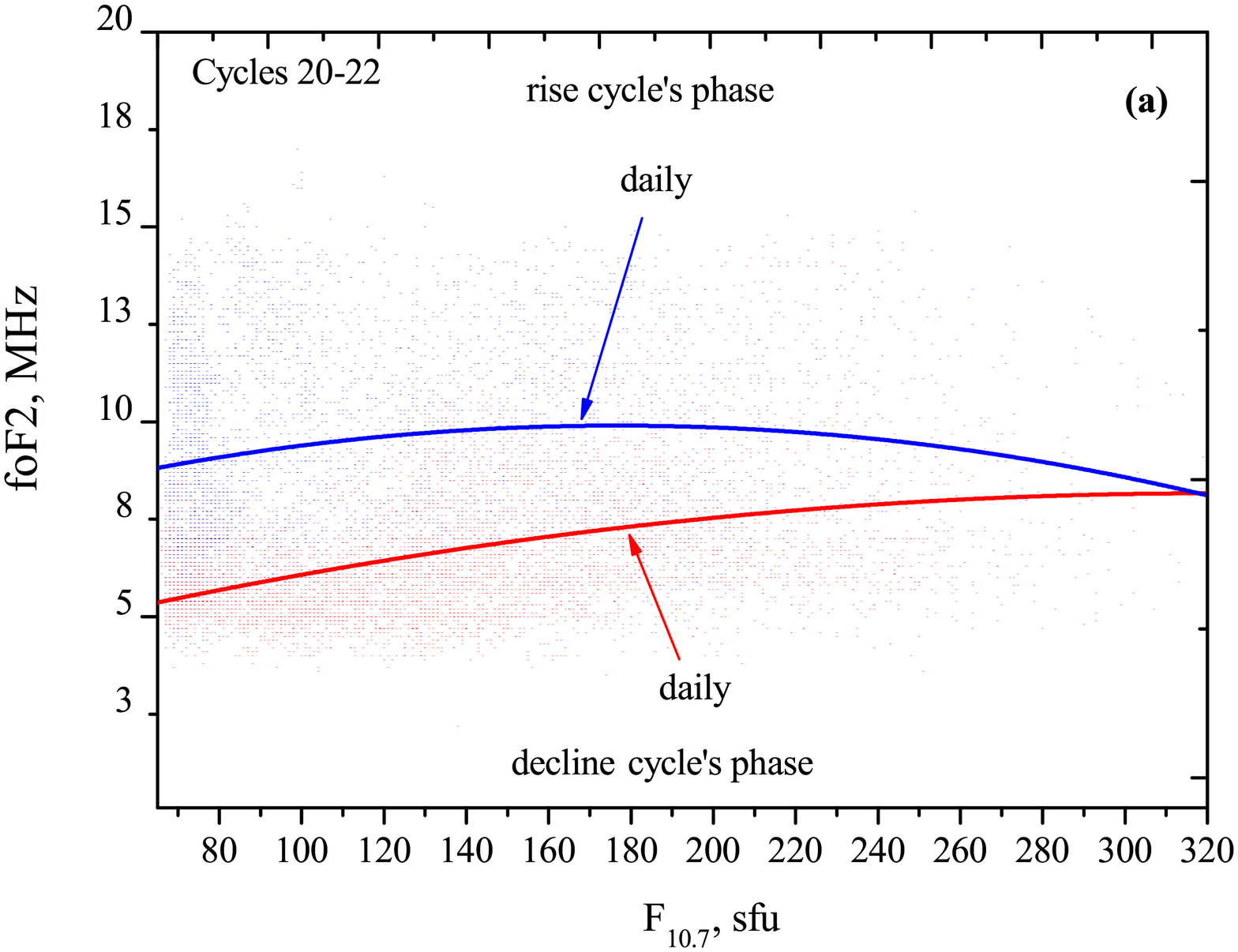}
               \includegraphics[width=70mm]{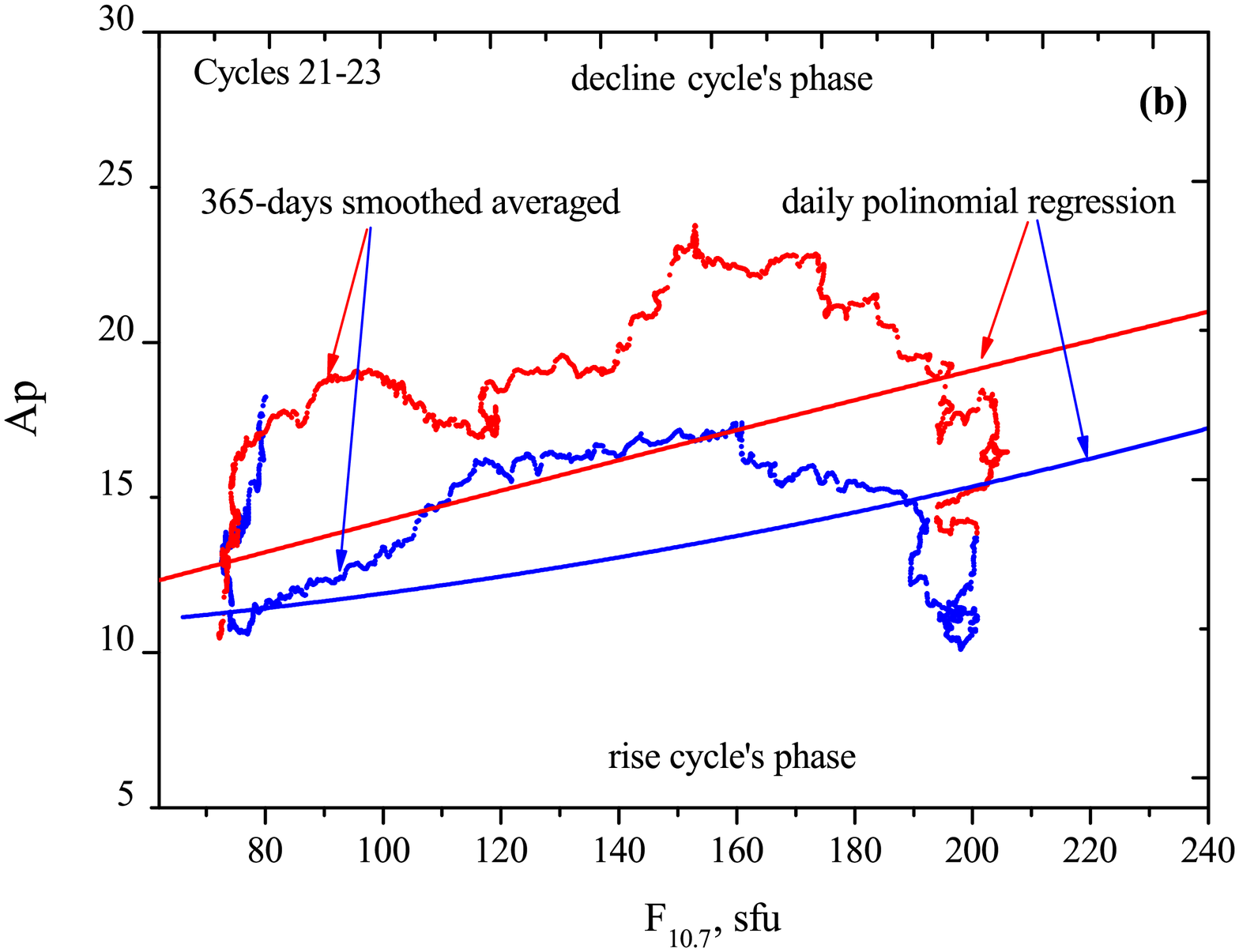}
              }

\caption{The daily values of  the ionospheric indices $foF2$ (a) and
of  the $A_p$ (b) depending on the daily averaged $F_{10.7}$}
 {\label{Fi:Fig6}}
\end{figure}

\begin{table}
\caption{The coefficients of the quadratic regression for daily
averaged values of the activity indices AI in the cycles 21, 22 and
23  in the phases of rise (R) and decline (D)}
\begin{tabular}{ccccccc}     
  \hline                   
AI $\leftrightarrow$ $F_{10.7}$             & A       & C1    & C2    & $\sigma_A$  & $\sigma_{C1}$& $\sigma_{C2}$    \\
SSN $\leftrightarrow$ $F_{10.7}$,R          & -81.2  & 1.35   & -9.4E-4 & 1.71       & 0.024        & 7.9E-5       \\
SSN $\leftrightarrow$ $F_{10.7}$,D          & 61.86  & -0.65  & 0.005   & 0.85       & 0.017        & 7.3E-5       \\
TSI $\leftrightarrow$ $F_{10.7}$,R          &1364.0  & 0.026  & -7.3E-5 & 0.04       & 5.52E-4      & 1.7E-6       \\
TSI$\leftrightarrow$ $F_{10.7}$,D           & 1364.7 & 0.013  & -1.7E-5 & -6.6E-4    & 9.5E-6       & 3.2E-8       \\
$F_{L \alpha}$$\leftrightarrow$ $F_{10.7}$,R& 2.21   & 0.022  &-2.7E-5  & 0.017      & 0.00025      & 9.2E-7       \\
$F_{L \alpha}$$\leftrightarrow$ $F_{10.7}$,D& 1.97   & 0.028  &-4.6E-5  & 0.02       & 0.00027      & 8.9E-7       \\
foF2 $\leftrightarrow$ $F_{10.7}$,R         & 7.19   & 0.031  &-8.7E-5  & 0.24       & 0.0037       & 1.3E-5       \\
foF2 $\leftrightarrow$ $F_{10.7}$,D         & 3.77   & 0.027  & -4.3E-5 & 0.16       & 0.0022       & 6.8E-6       \\
Ap      $\leftrightarrow$ $F_{10.7}$,R      & 10.24  & 0.008  &8.9E-5   & 1.21       & 0.002        & 6.3E-6       \\
Ap      $\leftrightarrow$ $F_{10.7}$,D      & 9.18   & 0.052  &-9.4E-6  & 1.32       & 0.018        & 2.2E-6       \\

\hline

\end{tabular}
\end{table}

It is known that the F2-region of ionosphere is formed under the
influence of ionizing solar EUV radiation in the thermosphere of the
Earth, which depends on the level of solar and geomagnetic activity.
One of the manifestations of the nonlinear relationship between the
foF2  the critical frequency of the F2 layer  and solar activity
indices  AI is the effect of the hysteresis of foF2 variations in
the 11 - yr cycle, [17].

Regular daily variations of the magnetic fields are created mainly
by changes of currents in the ionosphere due to the change of
illumination of the ionosphere by the Sun during the day. Irregular
variations of the magnetic field generated due to the flow of solar
plasma (solar wind) which is acted on the Earth's magnetosphere and
are associated with the flare activity of the Sun, changes in the
magnetosphere, and the interaction of the magnetosphere and
ionosphere. You must take into account that the ionospheric activity
indices dependent on the level of solar activity in a complex
manner: as in the case of the solar indices is the place ambiguous
dependence on $F_{10.7}$.

In Fig. 7 the solid lines show the regression line for the daily
data of the ionospheric indices fo2 and Ap. It is seen that
ionospheric indices weakly depend on the level of solar activity.
The effect of hysteresis on the  rise and decline cycle's phases is
expressed greatly. Fig. 7b presents a 365-day moving averages of Ap
index for 21 cycles. The effect of hysteresis in the case of the
365-day moving averages is more pronounced.

\section{The effects of hysteresis in the cyclic activity of stars in HK
project}

At the Mount Wilson Observatory in the framework of the HK project
the regular observations of the star - analogues of the Sun were
conducted from 1965 to the present. For the stars which were
selected for the HK-project program  it was defined by a
comprehensive index the Mount Wilson S-index (S-index Call) -
subsequently, the most important standard for characterizing the
activity of the atmospheres of stars in the optical spectral range,
see [4]. This program was specifically designed to study the cyclic
activity in F, G and K stars of the solar type. In the result of
long-term observations (over 40 years) it was significantly detected
the activity cycles for about 50 stars. It was found that on the
surface of stars, there are inhomogeneities that exist and vary over
multiple periods of rotation of the star around its axis. In
addition, the evolution of the rotation periods of the stars clearly
indicates the existence of differential rotations of the star,
similar to the differential rotations of the Sun, [4, 5]. The
evolution of active regions on the star on a time scale of about 10
years determines the cyclic activity similar to the Sun.

For 111 HK-project stars the, periodograms were computed for each
stellar record in order to search for activity cycles, [4]. The
significance of the height of the tallest peak of the periodogram
was estimated by the false alarm probability (FAP) function. The
stars with cycles  were classified as follows: if for the calculated
$P_{cyc} \pm \Delta P$ the FAP function $\leqslant 10^{-9}$ then
this star is of "Excellent" class ($P_{cyc} $ is the period of the
cycle). If $ 10^{-9} \leqslant FAP \leqslant 10^{-5} $ then this
star is of "Good" class. If $ 10^{-5} \leqslant FAP \leqslant
10^{-2} $ then this star is of "Fair" class. If $ 10^{-2} \leqslant
FAP \leqslant 10^{-1} $ then this star is of "Poor" class.

At the Lowell  and Fairborn Observatories the photometric
observations of the (Stromgren b, y photometry) was carried out in
parallel with observations of the S-index Call at the Observatory
Mount Wilson for the 32 stars from the same sample in a period of
13-20 years [5]. In most cases, the observed link between fluxes in
the chromospheres and photospheres of the stars: for one samle of
stars there is positive correlation between these fluxes (as in the
case of the Sun), for the others - the negative correlation.

In [18] it was made the analysis of the data on variability of
optical radiation  for the stars of the HK-project and on
variability of optical radiation the  stars  observed at the Crimean
astrophysical Observatory. The close relationship between spotting
of the stars with very different levels of activity and power of
their x-ray emission  was found [18,19]. The analysis of atmospheric
activity of solar-type stars using observations, including
HK-project and three broad program of searches for planets, in which
it were  determined the magnitudes of the  Call S-index for
thousands of stars showed that stars of the HK-project are the
closest to the Sun on the level of chromospheric, coronal radiation
and the cyclical activity [19, 20, 21].

We chose 4 star of the HK-project with the highest quality of
cyclical activity of "Excellent" class. This class corresponds to
class calculated for them according to values of FAP [Bal]. These
stars are: HD 103095 (a); HD 160346 (b), HD 81809 (C); HD 152391
(d). These stars belong to spectral classes G and K. For these stars
in [5] the graphs  of simultaneous observations chromospheric Mount
Wilson S-index and photometric observations at Lowell Observatory
are presented. The star HD 103095 and HD 160346 are characterized by
a cyclic activity with periods of about 7 years, we studied 3 full
cycle of activity. For the star HD 81809 and HD 152391 with longer
cycles we studied 2 full cycles of activity only. Using the
graphical dependencies from [4,5], we created a dataset for these 4
stars (which represent the average value for 3 months of
observations). So we have made a nesessary for our study the couple
of indices  of chromospheric and photospheric activity: S-index Call
versus Fphotosphere separately for the phases of rise and phases of
decline of star's activity cycles.

\begin{figure}[h!]
   \centerline{\hspace*{0.005\textwidth}
               \includegraphics[width=70mm]{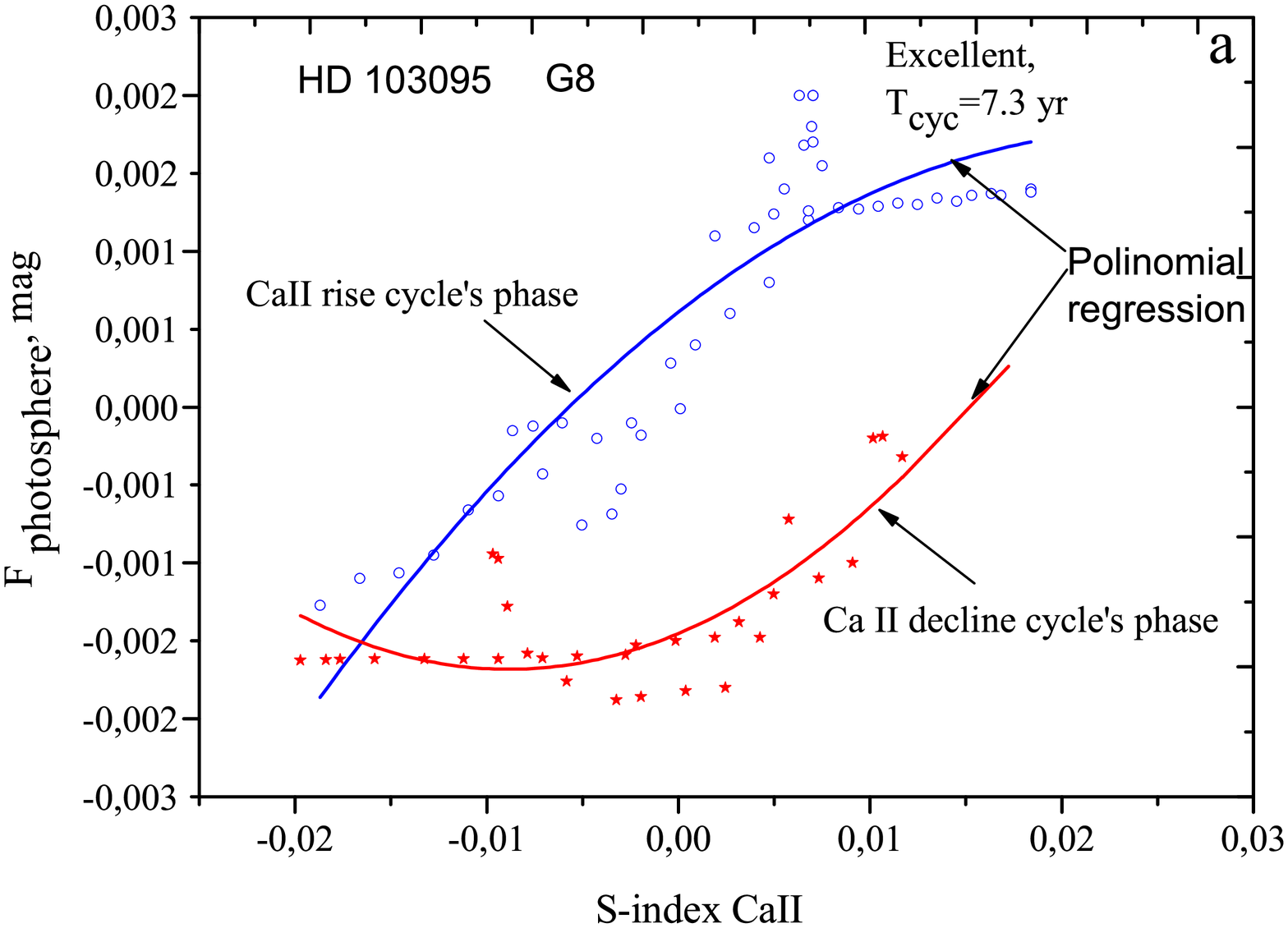}
               \includegraphics[width=70mm]{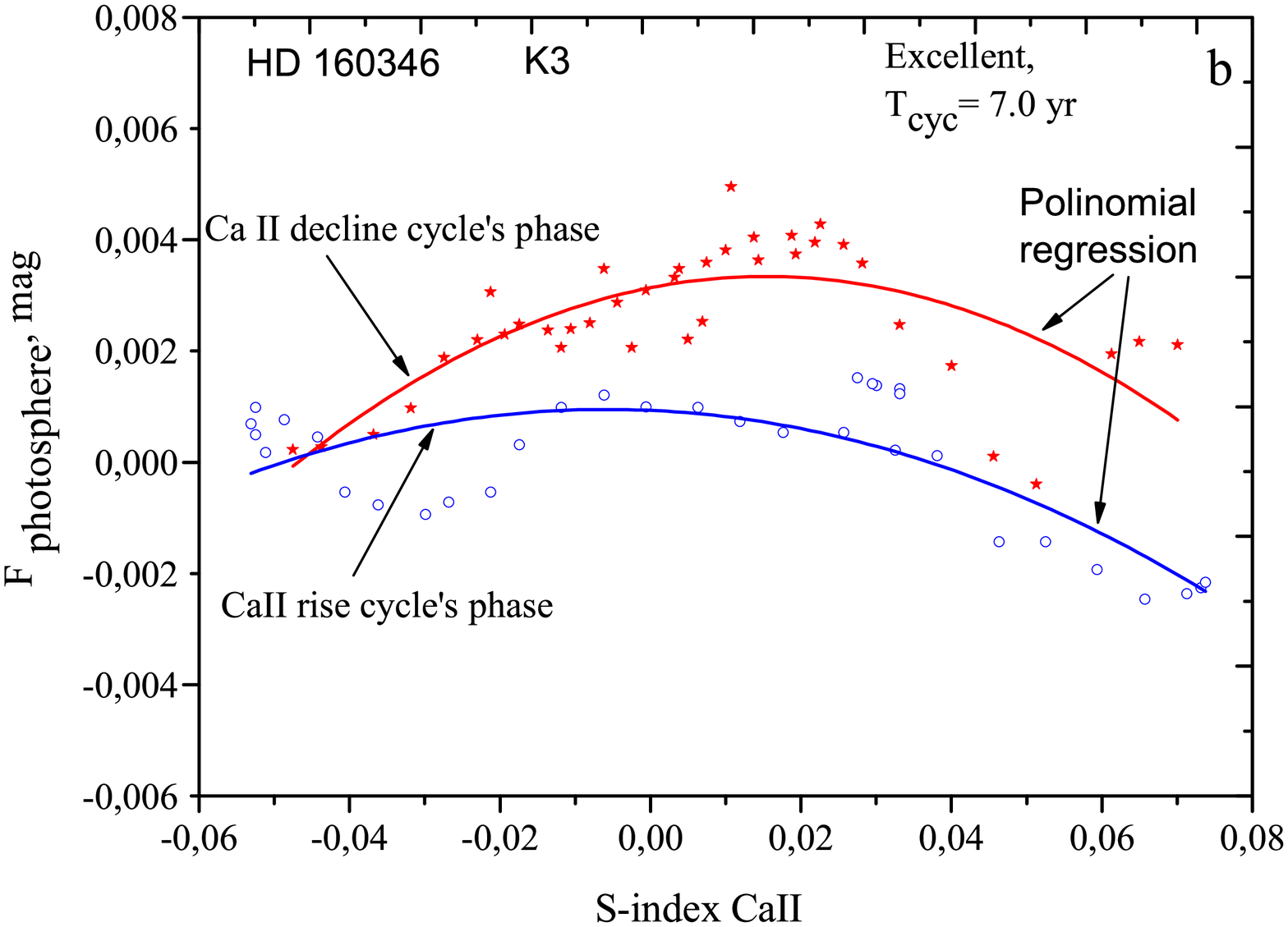}
              }
\centerline{\hspace*{0.005\textwidth}
               \includegraphics[width=70mm]{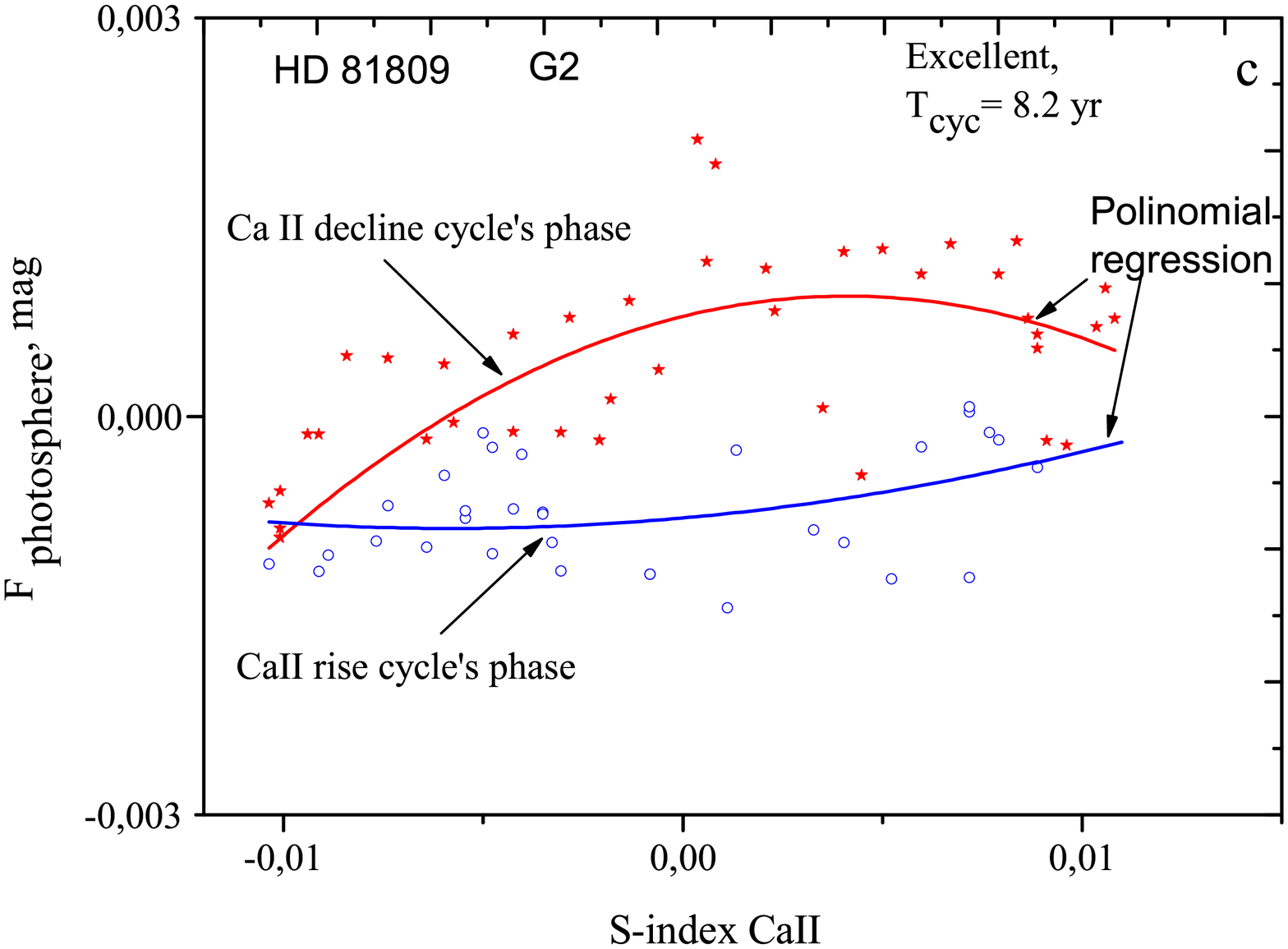}
               \includegraphics[width=70mm]{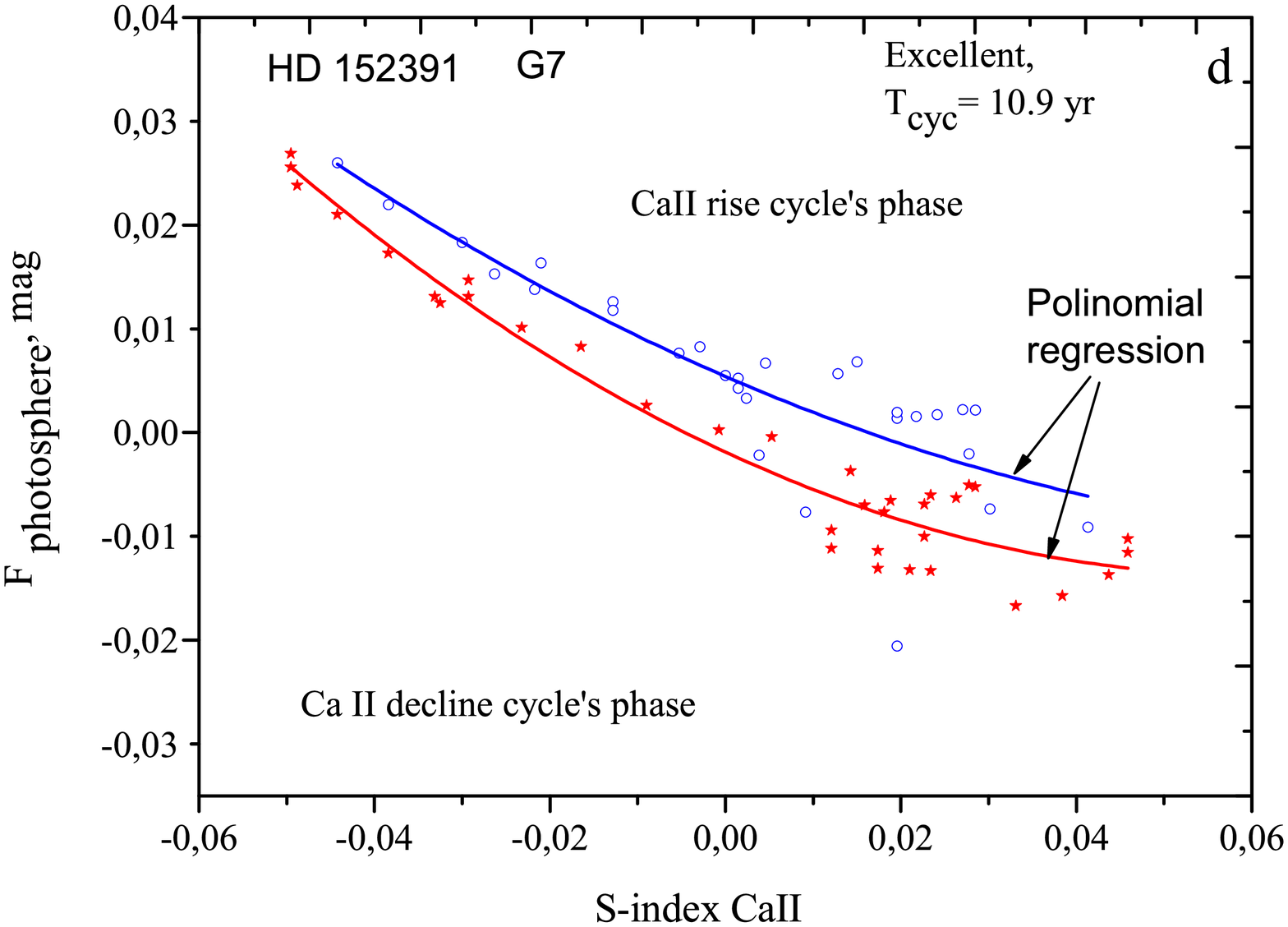}
              }
\caption{The effect of hysteresis in stars with activity cycles
between chromospheres S-index CaII and the flow of radiation from
the photosphere $F_{photosphere}$: HD 103095 (a); HD 160346 (b), HD
81809 (C); HD 152391 (d). The hollow circles correspond to the
phases of rise of the cycles for the S CaII index, the asterisks
correspond to the phases of decline of the cycles.}
 {\label{Fi:Fig8}}
\end{figure}

In Fig. 8 we present the averaged over 3 months of observations the
S-index Call values depending on the $F_{photosphere}$ values. Solid
lines indicate regression curves for the rise cycle's phase and
decline cycle's phase of stellar cycles chromospheres activity.

The data of observations of stellar indices are also being explored
with polynomials of second order. It is seen that for the selected
stars with the cyclicity of the class "Excellent" there is a
hysteresis effect for the analyzed pairs of indices characterizing
chromosphere and photospheric activity. Phases of rise and decline
in the cycles of activity are determined by increase and decrease of
a universal index of activity of the chromosphere, the S-index Call.
Note that with the increase in the activity of the chromosphere,
photosphere activity does not always increase (as in the case of the
Sun and stars HD 103095 and HD 81809 in Fig. 8a and 8c) and may
remain approximately constant or decrease, see Fig. 8b (HD 160346)
and 8d (HD 152391).

\section{Discussion}

We can see from the above consideration that the hysteresis effects
exist for the various manifestations of solar activity from
variations in the short wavelength range to the optical and radio
bands, and are found in the manifestations of ionospheric and
geomagnetic activity, and solar-type stars with detected cycles of
activity.

For individual activity indices  ($F_{L \alpha}$, foF2, $F_{10.7}$)
the effect of hysteresis can be explained using the two-component
model, [22]. According to the two-component model the radiation flux
$I_{\lambda}$ in the line is equal to:

\begin{equation}
    I_{\lambda} = B0+ B1 \cdot (F_B-60)^{2/3} + B2 \cdot (F_{10.7} - F_B)^{2/3}  \,.
   \end{equation}

where $F_B$ is the background radiation flux from the undisturbed
surface of the solar disk (with no active regions). Background
radiation $F_B$ on the wave of 10.7 cm varies in a cycle of solar
activity from 60 to 120 sfu, [23].

The analysis of cyclic variations in $F_{10.7}$ showed that, on
average, an empirical regression of the ratio of radio flux $
F_{10.7} = a + b \cdot F_B$ is performed. Here the coefficients a
and b characterize the contribution from the two components of the
model and change (1) at different phases of the activity cycles, and
(2) from cycle to cycle, [22].

This fact determines the uneven relative decline and the increase of
flows for the studied pairs of indices of activity at different
phases of the solar cycle, which leads to hysteresis. The flow in
weak cycles is determined mainly by the background radiation flux,
and in strong cycles are relatively more important becomes the
contribution from active regions of hysteresis effects should be
more pronounced in stronger cycles, see [22]. A possible mechanism
for the emergence of hysteresis at the present time is explained as
a result of the asymmetry of solar Dynamo in the different phases of
magnetic cycles, [24, 25].

\section{Conclusions}

The effect of hysteresis is typical not only for pairs of activity
indices in 11-yr solar cycles, but for the stars of the HK- project
with detected the stable cycles of activity similar to the Sun.
Hysteresis is a real delay in the onset of the maximum and decline
phase of solar and stellar activity and is an important key in the
search for physical processes responsible for changing radiation at
different wavelengths.

The effect of hysteresis is apparently a common property of
astronomical systems, which are characterized by different
manifestations of the cyclic activity associated with the time
evolution of magnetic fields. The hysteresis of foF2 is exist due to
the ionospheric response to variations in solar activity, in
particular, we can see the hysteresis in EUV radiation and
hysteresis of the geomagnetic activity index AP. All this hysteresis
effects is due to the hysteresis of the solar Dynamo.

For pairs of indices of solar activity and solar-type stars activity
the effect of hysteresis appears in different ways. The  curves of a
hysteresis on the rise and decline cycle's phases vary from one
cycle to another.

\end{document}